\pdfoutput=1
\documentclass[twocolumn,amsmath,amssymb,10pt,aps,superscriptaddress,prl]{revtex4-1}

\usepackage{hyperref}
\usepackage{graphicx}
\usepackage{color}

\begin{document}
\title{ Vestiges of Topological Phase Transitions in Kitaev Quantum Spin Liquids }
\author{Ara \surname{Go}}
\email[]{The first two authors contribute equally to this work.}
\affiliation{Center for Theoretical Physics of Complex Systems, Institute for Basic Science (IBS), Daejeon 34126, Korea}
\author{Jun \surname{Jung}}
\email[]{The first two authors contribute equally to this work.}
\affiliation{Department of Physics, KAIST, Daejeon 34141, Korea}
\author{Eun-Gook \surname{Moon}}
\email[]{egmoon@kaist.ac.kr}
\affiliation{Department of Physics, KAIST, Daejeon 34141, Korea}

\begin{abstract}  
	We investigate signatures of topological quantum phase transitions (TQPTs) between the $Z_2$ quantum spin liquids (QSLs).
	In two spatial dimensions,  $Z_2$ QSLs and their TQPTs are only well defined at zero temperature ($T=0$), and it is imperative to clarify their observable signatures under nonzero temperatures.
	Here, we present the vestiges of TQPTs between $Z_2$ QSLs with Majorana fermions in terms of thermal Hall conductivity $\kappa_{xy}$ at nonzero temperatures.
	The $\kappa_{xy}/T$ shows characteristic temperature dependences around TQPTs. 
	We argue that an exponential upturn near $T=0$ and the peak of $\kappa_{xy}/T$ around massive excitation energy are observable smoking-gun signals of the TQPTs.
	Quantum critical fan-shape temperature dependences are uncovered across TQPTs. 
	We also perform the parton mean-field analysis on a modified Kitaev model with next-nearest neighbor interactions finding TQPTs between the phases with different Chern numbers and their vestiges self-consistently.
	We discuss the implication of our results to the recent experiments in $\alpha$-RuCl$_3$.
\end{abstract}

\date{\today}

\maketitle

{\it Introduction.}
Quantum spin liquids (QSLs) intrinsically host enormous entanglements, manifested by emergent excitations, fractionalized particles, and gauge fluctuations \cite{Zhou, Balents}. 
Theoretical investigations, mostly using parton constructions \cite{Sachdev, WenBook}, deepen our understanding in characteristics of QSLs, and even the existence of $Z_2$ QSLs of spin $1/2$ systems on a honeycomb lattice is shown by an exactly solvable model, the so-called Kitaev model \cite{Kitaev}.  
The Kitaev representation of spins, $\vec{S}= i c  \vec{b}$ with four Majorana fermions ($c, \vec{b}$), is used to prove the existence of a topological phase with the Chern number $|\nu|=1$. In this Letter, $Z_2$ QSLs with Majorana fermions as elementary excitations are called Kitaev QSLs.

 Important theoretical advances in Kitaev QSLs were achieved by Jackeli and Khaliullin who showed that strongly spin-orbit coupled honeycomb lattices may host  the Kitaev model in 2D \cite{Jackeli}.
 Several materials with 4$d$ and 5$d$ orbital degrees of freedom including $\alpha$-RuCl$_3$ have been suggested as candidates \cite{Nasu2,Perkins,Trebst2017, KimHS}.
 Extensive experimental works have been reported in neutron, specific heat, nuclear magnetic resonance, magnetic torque, and thermal conductivity in the materials \cite{Nagler, Ji, Burch, NasuNMR, Baek, Buchner, LeeM, Kasahara, Klanjsek, Matsuda}. Especially, the recent thermal conductivity experiment on $\alpha$-RuCl$_3$ reported the quantized thermal Hall conductivity of $\kappa_{xy} / T = (\pi/12) (k_B^2/\hbar)$ below 5K, a hallmark of a Majorana edge mode \cite{Matsuda}.
 In the experiment, a peak was observed around 10K, where a gauge flux gap was found \cite{Baek}. Its origin remains as an intriguing open question. 

Topological properties of Kitaev QSLs are not well defined at nonzero temperatures and adiabatically connected to high temperature paramagnetic states. Since it is impossible to detect topological properties sharply at nonzero temperatures, it is quintessential to search for vestiges of topological properties such as the quantization behavior of $\kappa_{xy}/T$  at sufficiently low temperatures.    
In this Letter, we investigate vestiges of topological quantum phase transitions (TQPTs) in Kitaev QSLs by using path-integral and parton mean-field analysis. It is shown that characteristic temperature dependence of $\kappa_{xy}/T$ may appear around TQPTs, and  we propose them as smoking-gun signatures of TQPTs.

{\it Path-integral formalism.}   
The Kitaev representation gives a Hilbert space of Majorana fermions, whose dimension is larger than one of spins. 
To describe spin physics, it is crucial to project out unphysical states in the Majorana Hilbert space. 
One conventional way is to define  the projection operator, $\mathcal{P} = \prod_{j} (\frac{1+ b_j^x b_j^y b_j^z c_j}{2})$ with a site index $j$  \cite{Kitaev, Z2}, and a spin state is obtained by applying the operator to a Majorana state, $ |\Psi_\mathrm{spin} \rangle = \mathcal{P} |\Psi_\mathrm{Majorana} \rangle$. 
However, it may be subtle and difficult to employ the projection operator for calculations of physical quantities such as dynamic spin susceptibility \cite{knolle2014}. 
{\it A priori}, one should first check whether a Majorana state, which may be obtained by the parton mean field analysis, produces a well defined spin state.  
Below, we consider the path-integral formalism and find nonperturbative properties that allow us to circumvent a part of the subtleties and difficulties from the projection operator.   

We use the path-integral formalism with four Majorana fermions, and the partition function is  
\begin{eqnarray}
\mathcal{Z} &=& \int \mathcal{D}c \, \mathcal{D }\vec{b} \,\prod \delta(b^x b^y b^z c -1) \, e^{- \mathcal{S}}. \nonumber 
\end{eqnarray}
Remark that the product of the delta functions precisely describes the projection operator with the implicit space-time index. 
The action of the spin Hamiltonian $H_\mathrm{spin}(\{\vec{S}_j \})$  is
\begin{eqnarray}
\mathcal{S} &=& \int_0^{\beta} d \tau \sum_j   \big( c_j \partial_{\tau} c_j + \vec{b}_j \partial_{\tau} \vec{b}_j \big) + \int_0^{\beta} d \tau H_\mathrm{spin} (\{c \}, \{\vec{b}\} ). \nonumber
\end{eqnarray}
For a generic spin Hamiltonian, $H_\mathrm{spin} = \sum_{j,k,\alpha,\beta} J_{jk}^{\alpha \beta} S_j^{\alpha}S_k^{\beta}$, we introduce three Hubbard-Stratonovich (HS) fields ($\lambda_j, u_{jk}, v_{jk}$), which give 
$\mathcal{Z} =  \int \mathcal{D}c_j \, \mathcal{D }\vec{b}_j  \mathcal{D}\lambda_j \mathcal{D} u_{ij} \mathcal{D} v_{ij} \, e^{-\mathcal{S}_\mathrm{eff}}$.
The effective action is
\begin{eqnarray}
\mathcal{S}_\mathrm{eff} &=& \int_0^{\beta} d \tau \sum_{j}   c_j \partial_{\tau} c_j + \vec{b}_j \partial_{\tau} \vec{b}_j + \lambda_j (b_j^x b_j^y b_j^z c_j-1) \nonumber \\
&+&\int_0^{\beta} d \tau  \sum_{j,k} (- i) u_{jk} \, c_j c_k + (+i) v_{jk} J_{jk}^{\alpha \beta} \, b_j^{\alpha} b_k^{\beta}  - u_{jk} v_{jk}. \nonumber
\end{eqnarray}
The Lagrange multiplier field $\lambda$ is redefined to absorb the factor $i$ as usual \cite{Z2}. 
One advantage of the path-integral formalism is that the projection operator can be treated as the four-fermion interaction term with $\lambda$ in sharp contrast to other bilinear constraints of fermions and bosons. As shown below, the four-fermion interaction is more irrelevant to the zeroth-order stationary approximation of the path integral.  
We emphasize that  no approximation is made in $\mathcal{Z}$ and the projection or interaction terms are simply rewritten in terms of the HS fields ($u_{jk}, v_{jk}, \lambda_j$), which allow the nonperturbative analysis below. 

First, a quantum state without mixing $c$ and $\vec{b}$ Majorana fermions ($|{\Psi}_{M} \rangle \equiv  | \{c\} \rangle \otimes |\{\vec{b}\} \rangle $) describes a paramagnetic state because of the properties, $\langle \{c\}  |c_j | \{c\} \rangle = \langle \{\vec{b}\}  |\vec{b}_j | \{\vec{b}\} \rangle =  0$, which is our main focus in this Letter. The projection operator is spin singlet, so the corresponding spin state describes a QSL. 
In the path-integral formalism, the nonmixing condition indicates that the four-fermion interaction term ($b_j^x b_j^y b_j^z c_j$) is subdominant. If the interaction term is rewritten with an additional HS field, its mean value is zero for a QSL because of $\vec{S} = i c \vec{b}$. 
 
Second, for a gapped QSL, the standard stationary approximation to the partition function is safe. Namely, a  self-consistent mean-field solution is reliable, and the HS fields may be replaced by their mean values, $\bar{u}_{jk} \equiv (+i) \langle J_{jk}^{\alpha \beta} b^{\alpha}_j b^{\beta}_k \rangle$, $\bar{v}_{jk}=(-i)\langle c_j c_k \rangle $ omitting the bar notations hereafter. The four Majorana bands are well defined with the Chern numbers ($\nu_n$) and bulk energy gaps ($\Delta_n$) with a band index, $n=1,2,3,4$. The total Chern number is $\nu \equiv \sum_{n} (\vec{\nu})_n = \sum_n \nu_n$, and stability of a gapped QSL is guaranteed by the energy gaps. 
At a TQPT, an energy gap closes  (say, $\Delta_4 =0$), and if the gap closing happens only at few points in Brillouin zone with linear dispersion relations, the projection or interaction terms are irrelevant as in the $B$ phase of the Kitaev model \cite{Kitaev}. We consider such TQPTs in this Letter. 
The path-integral formalism demonstrates the existence of the gapped QSLs and their TQPTs with mean-field {\it Ans\"{a}tzs}. It also shows that the phases and their transitions are not destabilized by the four-point interaction term from the projection operator.

\begin{figure}
	\includegraphics[width=0.99\columnwidth]{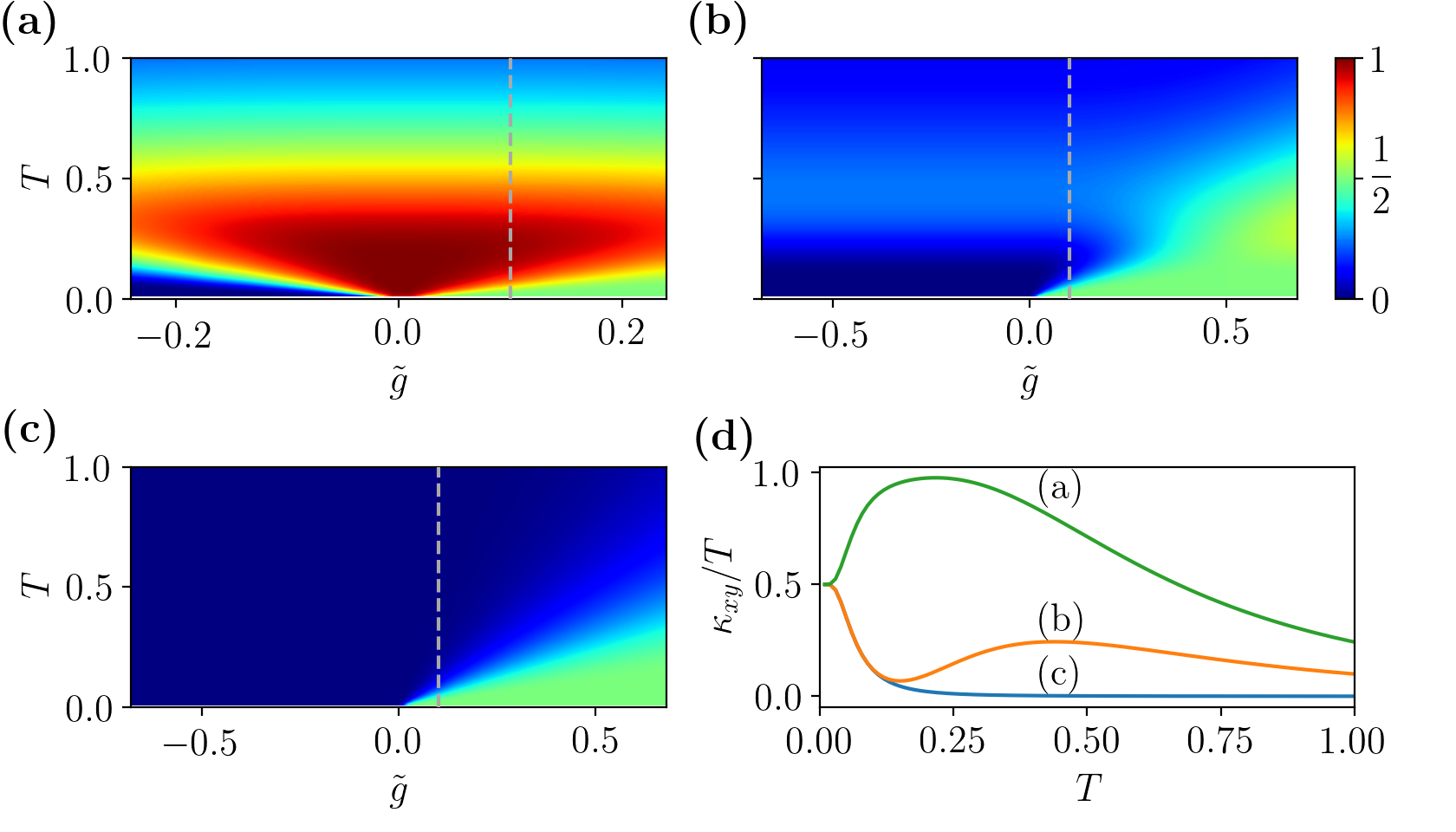}
	\caption{
	 { $\kappa_{xy}^\mathrm{edge}/T$ near TQPTs between the phases with $\nu=0$ and $\nu=1$. 	 
	 The units of $\kappa_{xy}^\mathrm{edge}/T$ and $T$ are $(\pi/6) (k_B^2/\hbar^2)$ and $\Delta_1/2$, respectively.	
	 The smallest gap depends on $ |\tilde{g}|$ linearly, ($\Delta_4= \Delta_1 |\tilde{g}| $). 
 $\tilde{g} > \tilde{g}_c$ is for the phases with $\nu=1$, and its false color representations  of $\kappa_{xy}^\mathrm{edge}/T$ are given in (a)-(c). 
	 TQPT between (a) $\vec{\nu}=(1, 1, 0, -2)$ and  $(1, 1, 0, -1)$ with $\Delta_1=2$ and $\Delta_2=1$,
         (b) $\vec{\nu}=(1, -1, 0, 0)$ and  $(1, -1, 0, 1)$ with $\Delta_1=2$ and $\Delta_2=1$, and
	 (c) $\vec{\nu}=(1, -1, 0, 0)$ and  $(1, -1, 0, 1)$ with $\Delta_1=\Delta_2=2$.
	 (d) $\kappa_{xy}^\mathrm{edge}/T$ at $\tilde{g}=0.1$ (marked by dashed lines in (a)-(c)).
	}
	}
	\label{fig:phase}
\end{figure}

Armed with the nonperturbative properties, an edge theory of a gapped Kitaev QSL is naturally constructed,
$ \mathcal{L}_\mathrm{edge} =  \psi_n \partial_{\tau} \psi_n + \psi_n \epsilon_n (-i \partial_x) \psi_n$, with an edge coordinate $x$, energy dispersion $\epsilon_n(k_x)$, and a real Grassman field $\psi_n$. The energy dispersion of a chiral mode is well defined in the interval $\epsilon_n(k_x) \in (0, \Delta_n)$ if $\nu_n \neq0$. Defining an edge thermal current, $J_e(T)= \sum_n \int \frac{d k_x}{2\pi} v_{n}(k_x) \epsilon_n(k_x) f[\epsilon_n(k_x)]$ with a velocity of the mode ($v_{n}$) and the Fermi distribution function ($f$), the thermal Hall conductivity divided by temperature is obtained, 
\begin{eqnarray}
	\frac{\kappa_{xy}^\mathrm{edge}}{T} = \sum_{n} \nu_{n}\big\{ \frac{\pi}{12} -\frac{1}{2\pi T^3}\int^{\infty}_{\Delta_{n}} \frac{\epsilon^2 e^{\epsilon/T}}{(1+e^{\epsilon/T})^2} d \epsilon \big\}. \label{edge}
\end{eqnarray}
We use the units ($ k_B=\hbar=1$) (see Supplemental Material~\cite{supp}). The edge current becomes exact when bulk bands are flat (localized).
Its quantization is obvious in the zero temperature limit ($\kappa_{xy}^\mathrm{edge}/{T} \rightarrow \nu \pi /12$). 

Near a TQPT, we introduce a parameter  $\tilde{g}$ to describe a gapped Kitaev QSL with $\nu$ ($\tilde{g} < \tilde{g}_c$) and the one with $\nu+1$ ($\tilde{g} > \tilde{g}_c$). A critical value $\tilde{g}_c$ is set to be zero hereafter. We assume the lowest energy gap has the dependence $\Delta_4 \propto |\tilde{g}|$ near the TQPT and the corresponding mode has a linear dispersion relation, which is supported by our parton mean-field analysis below. Generalization to generic TQPTs is straightforward.  At the quantum critical point, using the edge theory Eq.~(1) may be subtle, but as shown below, the edge theory results are well matched with the bulk theory calculations.   
Ignoring $\tilde{g}$ dependences of the bigger energy gaps ($\Delta_{1,2,3}$), we illustrate ${\kappa_{xy}^\mathrm{edge}}/{T} $ for TQPTs between $\nu=0$ and $\nu=1$ in Fig.~1. Striking temperature dependences appear at nonzero temperatures, characterized by the structures of ($\nu_n, \Delta_n$), as shown in Figs~1(a)--1(c). Detailed conditions of ($\nu_n, \Delta_n$) are explained in the caption.

The critical-fan shapes manifest around  $(\tilde{g}_c =0, T=0)$ in Fig.~1(a), and the fan-shapes appear in $\tilde{g}_c >0$ in Fig.~1(b) and (c).
We present the characteristic curves for the $\nu=1$  phases [along the $\tilde{g}=0.1$ line in Figs.~1(a)--1(c]) in Fig.~1(d).
At $\tilde{g}=0.1$ in (a), we find 
\begin{eqnarray}
\frac{\kappa^\mathrm{edge}_{xy}}{T} \simeq \frac{\pi}{12} + \alpha_1 (\frac{\Delta_4}{T})^2 e^{-\frac{\Delta_4}{T}}, \quad \alpha_1 >0,
\label{eq:kappa}
\end{eqnarray}
in the limit of $T \ll \Delta_4$ with the conditions $\Delta_1=2\Delta_2 \gg \Delta_4$ and $\vec{\nu} = (1,1,-1,0)$.

The temperature dependence of the upturn is exponential, and the peak of $\kappa_{xy}/T$ appears around $T \sim \Delta_4$
where its height depends on $\nu_{1,2,3}$ and $\Delta_{1,2,3}$.
We propose the existence of the peak with the exponential upturn as a smoking-gun signature of a TQPT between $\nu=0$ and $\nu=1$.
On the other hand, the cases (b) and (c) show 
$ {\kappa^\mathrm{edge}_{xy}}/{T} = {\pi}/{12} - \alpha_2 ({\Delta_4}/{T})^2 e^{-{\Delta_4}/{T}}$,
with a positive constant $\alpha_2$. 
Such temperature dependences are similar to the ones of the original Kitaev model under weak magnetic field \cite{NasuThermal} but are significantly different from the ones in experiments \cite{Matsuda}. 

The qualitative differences between (a), (b), and (c) are originated from the overall structure of $(\Delta_n, \nu_n)$. Though the total Chern number is only important near $T=0$, all the Chern numbers and band gaps become relevant at nonzero temperatures. Thus, not only the matter Majorana fermions ($c$) but also the gauge-flux Majorana fermions ($\vec{b}$) can contribute to thermal Hall conductivity and it is crucial to keep both of them at nonzero temperatures.

\begin{figure}
	\includegraphics[width=2.8in]{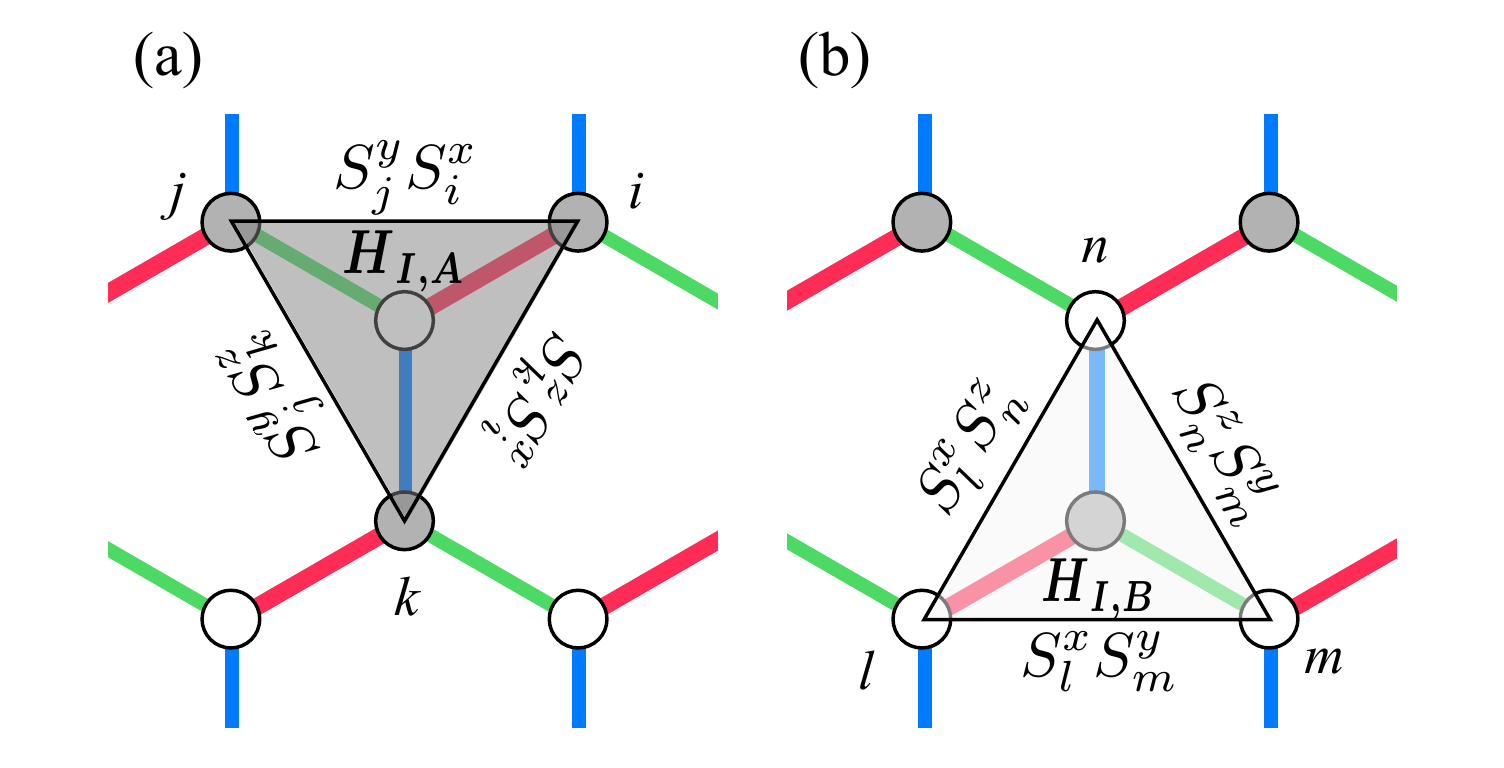}
	\caption{  {Graphical representation of the NNN spin exchange interactions of sublattice (a) $A$ and (b) $B$.
		Sites of different sublattices are illustrated by white and gray circles.  The nearest neighbor bonds that correspond to $S^x S^x$, $S^y S^y$ and $S^z S^z$ in the Kitaev Hamiltonian are plotted by red, green, and blue lines respectively.} }
	\label{fig:Lattice}
\end{figure}

  {\it Parton mean-field analysis.} To understand TQPTs with a microscopic perspective, let us consider a modified Kitaev Hamiltonian, $ H_\mathrm{tot} =  H_K+ g(H_{I,A} +H_{I,B})$, with  a dimensionless parameter ($g$). We find that the next-nearest neighbor (NNN) interaction terms, 
\begin{eqnarray}
H_{I,A} &=& - \frac{K}{2} \sum_{\bigtriangledown, ijk} S_i^x S_j^y + S_j^y S_k^z + S_k^z S_i^x \nonumber \\
H_{I,B} &=& - \frac{K}{2} \sum_{\triangle, lmn} S_l^x S_m^y + S_m^y S_n^z + S_n^z S_l^x,\nonumber
\end{eqnarray} 
capture TQPTs nicely. 
The summations over $\bigtriangledown, \triangle$ are graphically represented in Fig.~\ref{fig:Lattice}.
The original Kitaev Hamiltonian $H_K = -K \sum_{\langle i,j \rangle} S_{j}^{\alpha_{jk}} S_{k}^{\alpha_{jk}}$ is used whose link dependent exchange interactions with $\alpha_{jk}$ are given in Ref.~\cite{Kitaev}. For simplicity, we consider the isotropic exchange interactions, and its generalization to anisotropic cases is straightforward. 
We note that the NNN term respects all the symmetries of the Kitaev model including the time-reversal symmetry. Below, we show that by increasing the coupling strength $g$, the time reversal symmetry is spontaneously broken, which effectively plays a role of an applied magnetic field. 

%
%

For $g=0$, the {\it Ans\"{a}tz}, $u_0 = \left\langle {i} \, c^{}_j c^{}_k \right\rangle$ and $u = \left\langle {i} \, b_j^{\alpha_{jk}} b_k^{\alpha_{jk}} \right\rangle$ with the nearest neighbor indices ($j,k$) is used, 
and the mean-field Hamiltonian is 
\begin{eqnarray}
	H_K^\text{MF} = -K \sum_{\langle j,k \rangle} \left[u \left({i} c^{}_j c^{}_k \right) + u_0 \left({i} b_j^{\alpha_{jk}} b_k^{\alpha_{jk}}\right) - u_0 u\right]. \nonumber
\label{eqn:KitaevMF}
\end{eqnarray}
We find ($u_0 = - 0.5249$, $u = + 1$) giving the gapless excitations of $c$ Majorana fermions and the flat bands of $\vec{b}$ Majorana fermions, which are precisely the same as the previous literature \cite{You2012}.

\begin{figure}
\includegraphics[width=0.99\columnwidth]{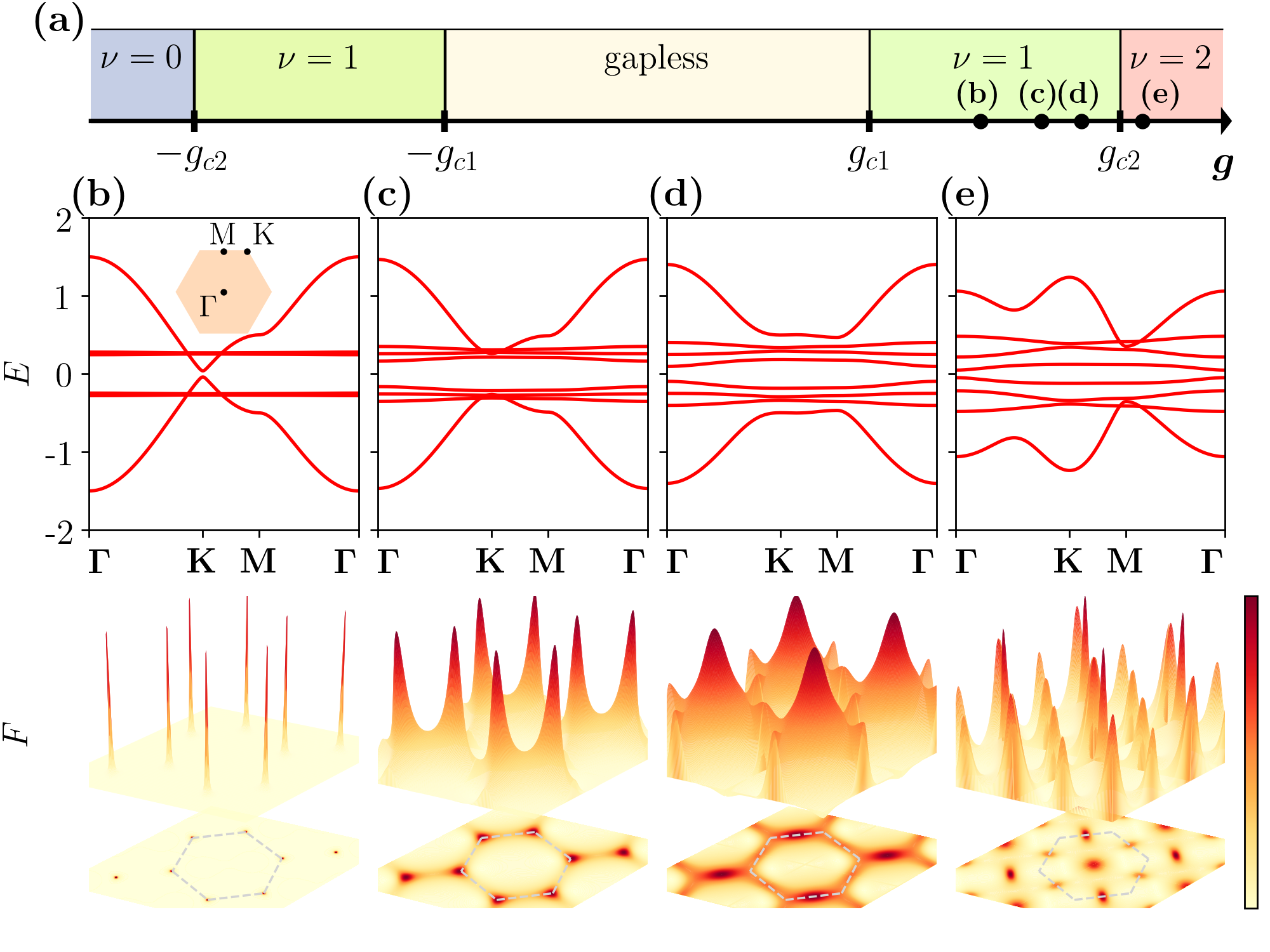}
\caption{  (a) Phase diagram of $H_\mathrm{tot} = H_K + g(H_{I, A} +H_{I, B})$ with parton mean-field analysis.
  Band structure along the high symmetry line for few selected values of $g$, (b) $g$=1.6, (c) $g$=1.9, (d) $g$=2.1, and (e) $g$=2.3.
The corresponding Berry curvature in the momentum space are given below the band structure.
The first Brillouin zone is marked by dashed lines in the projection.}
 \label{fig:PhaseDiagram}
\end{figure}

For $g\neq0$, we extend our mean-field analysis by introducing mean-values for the NNN interaction terms, $w_0 = \langle {i}\, c^{}_{j}\, c^{}_{k} \rangle$ and $w =  \langle {i} \, b_{l}^x \, b_{m}^y \rangle$ with the NNN indices ($j,k,l,m$). 
The mean-field {\it Ans\"{a}tz} gives
\begin{align}
\begin{split}
	H_{I,A}^\text{MF} = \frac{K}{2} \sum_{\bigtriangledown, ijk} &w\left({i}\,c^{}_{A,i} c^{}_{A,j}\right) - w_0 \left({i}\, b_{A,i}^x b_{A,j}^y \right) \nonumber\\
&+ w_0 w +\left[(x,y,z)\text{ cyclic terms}\right],  \nonumber
\end{split}\\
\begin{split}
	H_{I,B}^\text{MF} = \frac{K}{2} \sum_{\triangle, lmn} &w\left({i}\,c^{}_{B,l} c^{}_{B,m}\right) - w_0 \left({i}\, b_{B,l}^x b_{B,m}^y \right)\nonumber\\
&+ w_0w +\left[(x,y,z)\text{ cyclic terms}\right], \nonumber
\end{split}
\end{align}
where the sublattice indices ($A,B$) are shown explicitly for clarity. 
The summation over the triangle and reverse-triangle is done in the clock- and counterclockwise directions. 
Nonzero values of $w$ and $w_0$ break time-reversal symmetry, and their form is equivalent to the effects of a weak external magnetic field along the $(1,1,1)$ direction in perturbative analysis \cite{Kitaev}. In this sense, our parton mean-field analysis captures physics of the original Kitaev model under an external magnetic field by using spontaneous symmetry breaking.  For simplicity, we keep a threefold rotational symmetry, but weak symmetry breaking effects do not change our conclusions since topological phases are gapped. 

We find three different phases separated by {$g_{c_1}\sim 1.05$ and $g_{c_2} \sim 2.25$} for $g >0$. For $g <0$, only the Chern numbers of $\vec{b}$ Majorana fermions are opposite to the ones at $|g|$.   

(1) $ g  < g_{c_1}$, the {\it Ans\"{a}tz} and the band structures are the same as the ones of $g=0$. Time reversal symmetry is not broken.

(2) $ g_{c_1}<  g  < g_{c_2}$, all the mean-values are nonzero breaking time reversal symmetry. The typical band structures and corresponding Berry curvature~\cite{Xiao2010} $F(\mathbf{k})$ are illustrated in Fig.~\ref{fig:PhaseDiagram}. Clearly, all the bands are gapped and dispersive. 
We find  the Chern numbers of $\vec{b}$ Majorana fermions ($-1,0,1$), respectively. One way to understand the symmetry breaking may be related to energy gap opening of all Majorana fermions by breaking the time-reversal symmetry. We consider all available bilinear {\it Ans\"{a}tz} together and find that only the time reversal symmetry breaking one has the nonzero values. 

(3) $ g_{c_2}<g$,  we find a topological phase with $\nu=2$.  At $g=g_{c_2}$, one of the $\vec{b}$ Majorana bands becomes gapless while the $c$ Majorana fermions remain gapped, and the gapless excitation appears at $\Gamma$ with a linear dispersion relation.  
Our results self-consistently show the two different channels of TQPTs as in the previous analysis with vortex lattices \cite{Pachos2}.

\begin{figure}
	\includegraphics[width=3.2in]{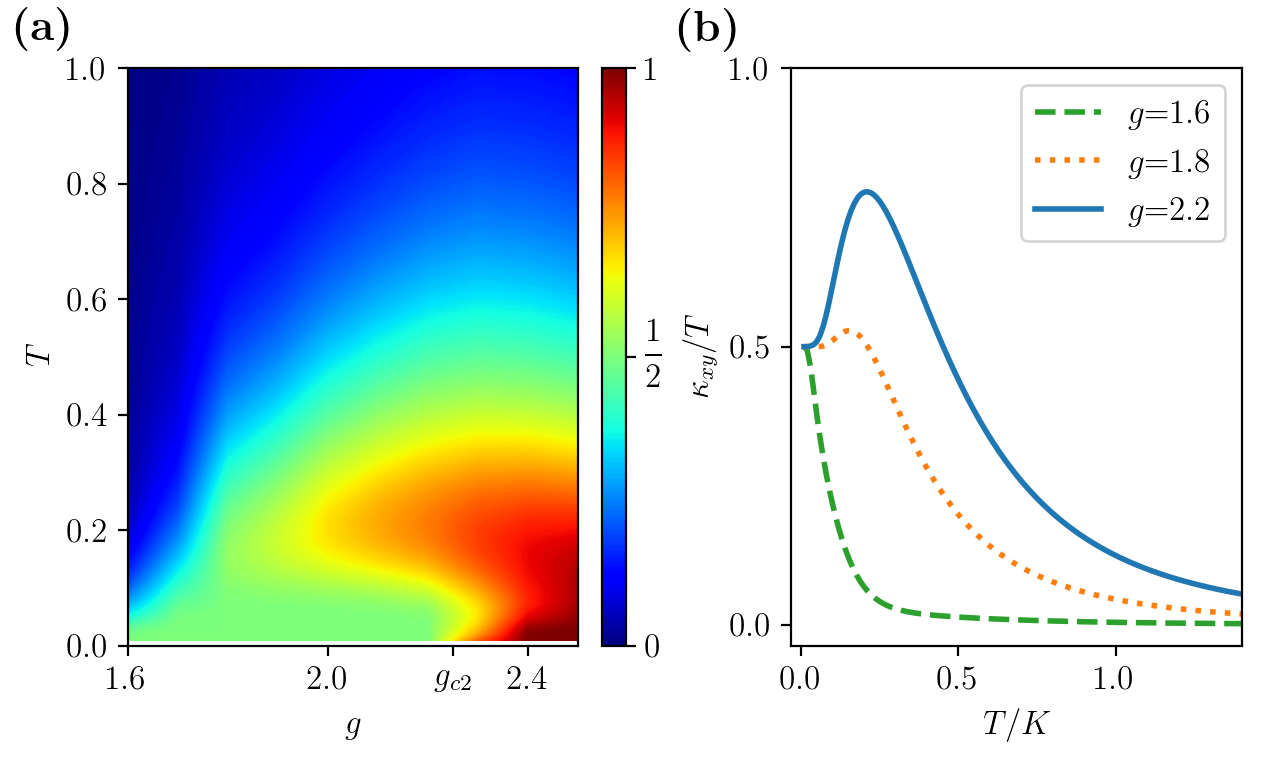}
	\caption{ (a) False color representation of $\kappa_{xy}/T$ on $g-T$ plane with $g_{c2} \sim2.25$. 
	(b) $T$ dependence of $\kappa_{xy}/T$  for $g$= 1.6 (dashed), 1.8 (dotted), 2.2 (plain). The unit of $\kappa_{xy}/T$ is the same as in Fig.~1. }
	\label{fig:ThermalConductivity}
\end{figure}

The bulk thermal Hall conductivity is computed by using the conventional formula~\cite{Vafek2001, Sumiyoshi2013}
\begin{align}
	\frac{\kappa_{xy}^\mathrm{bulk}}{T} = -\frac{1}{T^2} \int d \epsilon~ \epsilon^2 \frac{ \partial f (\epsilon, T)}{\partial \epsilon} \sum_{\mathbf{k},\varepsilon_n(\mathbf{k})<\epsilon} F_n(\mathbf{k}),
\end{align}
where the $\mathbf{k}$ summation runs over the first Brillouin zone and $\varepsilon_n(\mathbf{k})$ is the $n$th energy eigenvalue of the mean-field Hamiltonian. 
At $T\rightarrow 0$, ${\kappa^\mathrm{bulk}_{xy}}/{T}$ is quantized as $\nu{\pi}/{12}$ same as Eq.~(1). Our calculations are asymptotically exact at low temperatures and naturally become uncontrolled at temperatures higher than the gap energy scale.  

In Fig.~4(a), the density plot of ${\kappa^\mathrm{bulk}_{xy}}/{T}$ is illustrated. Qualitative dependences on  $T$ and $g$ are similar to Fig.~1. For example, the critical-fan shape appears as in Figs.~1(b) and 1(c) near $g_{c2}$, but away from $g_{c2}$, it is more distorted than Fig.~1 because higher energy bands depend on $g$ and have dispersive spectrums.  
In Fig.~4(b),  we plot temperature dependences of $\kappa^\mathrm{bulk}_{xy}/T$ for different values of $g$. For $g=1.6$, $\kappa^\mathrm{bulk}_{xy}/T$ monotonically decreases as the temperature rises, but for $g=2.2$, nonmonotonic temperature dependences appear with a peak around $T/K \sim \Delta_4$ and, for $T \ll \Delta_4$,  the upturn temperature dependence is exponential. The bigger band gaps $\Delta_{1,2}$ are in an order of magnitude larger than $\Delta_4$, and the Chern number vector is  $\vec{\nu}= (1,1,0,-1)$. The qualitative matching between $\kappa^\mathrm{edge}_{xy}/T$ and $\kappa^\mathrm{bulk}_{xy}/T$ is one sanity check of our analysis.

{\it Discussion and conclusion.} 
Our calculations of $\kappa_{xy}/T$ illustrate vestiges of TQPTs at nonzero temperatures. 
We note that, in experiments, $\kappa_{xy}/T$ includes phonon contributions, and recent theoretical works reported that  the half quantized value may be stable under the presence of phonon contributions \cite{Balents2,Rosch}. In other words, the two contributions may be added up for the total conductivity, $\kappa_{xy, \mathrm{tot}} \simeq \kappa_{xy, \mathrm{spin}} +\kappa_{xy, \mathrm{phonon}} $.
The phonon contributions are estimated to show power-law temperature dependent corrections, $\kappa_{xy, \mathrm{phonon}}/T \sim T^2$, to the quantized value, which was applied to explain  the upturn of $\kappa_{xy}/T$ in $\alpha$-RuCl$_3$ \cite{Rosch}.  

We, instead, propose the vestiges of a TQPT as an alternative route to explain the upturn of $\kappa_{xy}/T$. Our analysis shows that spin degrees of freedom near TQPT in Kitaev QSLs may exhibit the upturn with the half-quantization, and its exponential temperature dependence is one of key differences from phonon contributions. 
 In reality, both of the two contributions may be present in addition to impurities, and further detailed analysis of  $\kappa_{xy}/T$ is desired to determine dominant contributions.  For example, investigating applied magnetic field effects in Kitaev QSLs may provide information of dominant carriers of $\kappa_{xy}$.  

We stress that all Majorana fermions ($c, \vec{b}$) may contribute to physical quantities near generic TQPTs. This is in drastic contrast to the original Kitaev model with weak perturbations where only $c$ Majorana fermions are important at low temperatures. Our calculations with the edge and the parton mean-field theories naturally capture the contributions of all the Majorana fermions, and $\kappa_{xy}^\mathrm{edge}/T$ is {\it asymptotically exact} at low temperatures. The exponential upturn followed by the peak of $\kappa_{xy}/T$ is a concrete prediction of our results, which makes quantum critical-fan shape dependence around TQPTs reliable. 
At topological quantum critical points, additional scattering mechanisms between gapless modes could appear which might be important.  Future works that consider the scattering mechanisms near TQPTs with acoustic phonons are desirable. 

In this Letter, we use $\kappa_{xy}/T$ for vestiges of TQPTs since it is directly related to the Chern number in $T \rightarrow 0$, but other physical quantities such as specific heat, nuclear magnetic resonances, and neutron experiments would also show signatures of TQPTs. Further research on the physical quantities with a perspective of TQPTs  would be highly desired.  
Also, numerical and theoretical works on realistic magnetic Hamiltonians including Heisenberg, $\Gamma$ interactions, and magnetic fields would be important \cite{Kim_yb, Motome2, Hickey2019}. 
We emphasize that our analysis can be readily extended beyond the scope of the Kitaev QSLs. The phenomenological nature with ($\Delta_n, \nu_n$) should be applicable to generic $Z_2$ QSLs as well as conventional phases with topologically nontrivial structures, for example spin-wave bands \cite{Moore, Lu2018}.  Even in generic Chern insulators, our methods may be directly applied while the half-quantized value of $\kappa_{xy}/T$ is absent.

In conclusion, we study the vestiges of TQPTs in Kitaev QSLs by using path-integral and parton mean-field analysis.  Around TQPTs, characteristic temperature dependences of $\kappa_{xy}/T$ are obtained, including quantum-critical fan-shape dependences, and we provide  smoking-gun signatures of TQPTs which may be tested in future experiments.

 We thank E. Berg, L. Janssen, Y. B. Kim, Y. Matsuda, Y. Motome, N. Perkins, and A. Rosch for invaluable discussions and comments. We are especially grateful to Y. Matsuda for sharing unpublished thermal Hall data. EGM is grateful to A. Furusaki, Y. Mastuda and Y. Motome  for their hospitalities during the visits to RIKEN, Kyoto University, and  University of Tokyo.  
This work was supported by NRF of Korea under Grant No. 2017R1C1B2009176 (JJ, EGM), the POSCO Science Fellowship of POSCO TJ Park Foundation (EGM), and
Institute for Basic Science (IBS) in Korea under Grant No. IBS-R024-D1 (AG).

\bibliographystyle{apsrev4-1}
\bibliography{Kitaev}

\end{document}


\title{ Vestiges of Topological Phase Transitions in Kitaev Quantum Spin Liquids: Supplementary Information }

\maketitle
\beginsupplement

\section{Comments on the path integral formalism}
In this section, we comment a few advantages of the path integral formalism. 
As discussed in literatures \cite{Kitaev}, the Majorana representation of quantum spins requires the local constraint, 
\begin{eqnarray}
\hat{b}_x \hat{b}_y \hat{b}_z \hat{c} =1. \nonumber
\end{eqnarray}   
The local constraint should be considered carefully because it is practically non-trivial in the conventional analysis. For example, the conventional mean-field or Hartree-Fock analysis is not guaranteed to be valid a priori even with a large order parameter because the constraint acts on the Hilbert space.
On the other hand,  the path-integral formalism naturally implements the local constraint by introducing an auxiliary field, or equivalently Lagrangian multiplier field, and the stationary method of the path integral gives the reliable zeroth order approximation. 

Furthermore, it is explicitly shown that the local constraint introduces an interaction term with the four Majorana fermions at each site. In drastic contrast to the standard fermionic and bosonic spinon representations where their local constraints have two fermion or boson operators, the local constraint is more irrelevant in terms of the renormalization group sense. 
For example, even in the gapless B phase of the pure Kitaev model, the four-point interaction is irrelevant because only $c$ band is gapless and the others are gapped. Along with this line, the gapped quantum spin liquid is also stable under the local constraint since all the excitations are gapped, and one can argue that the local constraint is irrelevant to the ground state. 

The path-integral formalism, as usual, is also useful for understanding non-perturbative properties of Kitaev quantum spin liquids. 
 For example, one may introduce a Hubbard-Stratonovich field to consider a specific interaction in the action, 
\begin{eqnarray}
\sum_i \int d \tau \mathcal{B}^{\alpha} (i, \tau) (i c(i, \tau) b_{\alpha}(i, \tau)). \nonumber
\end{eqnarray}
The term may be obtained from the local constraint term and the definition of the spin operator, $S_{\alpha} = i c b_{\alpha} =i \epsilon_{\alpha \beta \gamma} b_{\beta} b_{\gamma}$. 
If the Hubbard-Stratonovich field $\langle \mathcal{B}^{\alpha} \rangle \neq 0$, the effective action contains the term of $i c(i) b_{\alpha}(i)  $, which mixes wave functions of $c$ and $b_{\alpha}$ Majorana fermions. By using symmetry analysis, it is obvious that the state with $\langle \mathcal{B}^{\alpha} \rangle \neq 0$ breaks spin rotational symmetry. Therefore, the state with decoupled wave functions of $c$ and $b_{\alpha}$ Majorana fermions describes a paramagnetic state.

\begin{figure}[btp]
	\includegraphics[width=0.99\columnwidth]{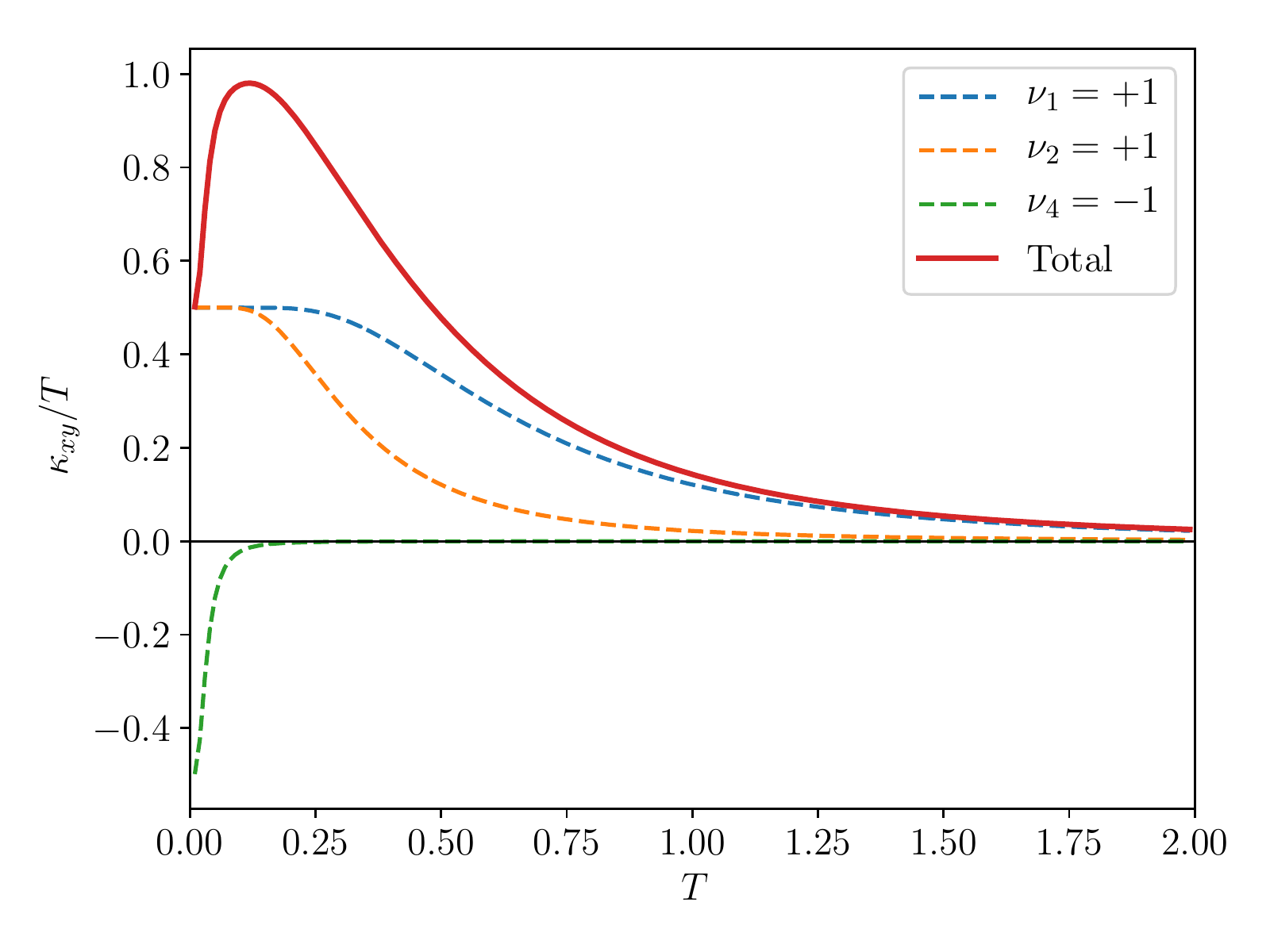}
	\caption{ $\kappa_{xy}/T$ from Majorana fermions. The red thick line is for the total $\kappa_{xy}/T$.  The blue (orange) dashed line is for the contribution of $\nu_1=1$ ($\nu_2=1$) and $\Delta=2$ ($\Delta=1$) to $\kappa_{xy}/T$. The green dashed line is for  the contribution of $\nu_4=-1$ and $\Delta=0.1$ to $\kappa_{xy}/T$. The different gap energy scales with the Chern numbers determine the peak position and height. 
	}
	\label{fig:kappa}
\end{figure}

 \section{Edge theory of thermal Hall conductivity}
To be self-contained, we present the derivation of $\kappa_{xy}/T$ by using the edge theory. We follow the notation of Ref.~\onlinecite{Rosch}.
 The temperatures along the top and bottom edges, $T_{top}$ and $T_{bot}$, are constant and the total energy current through the system is given by $J_T =J_e(T_{top})-J_e(T_{bot})$. For small temperature difference, the thermal current and the thermal Hall conductivity are 
 \begin{eqnarray}
 J_T = \frac{d J_e(T)}{dT} (T_{top}-T_{bot}), \quad \kappa_{xy} = \frac{d J_e(T)}{dT}. \nonumber
 \end{eqnarray}
 For a single chiral fermionic channel with arbitrary dispersion $\epsilon_{k_x}$, one finds
 \begin{eqnarray}
 \frac{d J_e(T)}{dT} = \int \frac{d k_x}{2\pi} \epsilon_{k_x} v_{k_x} \frac{d f (\epsilon_{k_x})}{dT}= - \int_{\epsilon_{min}}^{\epsilon_{max}} \frac{\epsilon^2 f'(\epsilon)}{2\pi T}  d\epsilon, \nonumber
 \end{eqnarray}
 with $v_{k_x} = \frac{d \epsilon_{k_x}}{d k_x}$. The chiral edge mode energy cutoffs are $\epsilon_{min}=0$ and $\epsilon_{max}= \Delta$ where $\Delta$ is a bulk energy gap. Thus, we find the formulas, 
 \begin{eqnarray}
  \frac{d J_e(T)}{dT} =  \int_{0}^{\Delta} \frac{\epsilon^2 }{2\pi T^2} \frac{e^{\epsilon/T}}{(1+e^{\epsilon/T})^2}  d\epsilon
 \end{eqnarray}
 and 
 \begin{eqnarray}
  \frac{\kappa_{xy}}{T}=  \frac{\pi}{12} -  \int^{\infty}_{\Delta} \frac{\epsilon^2 }{2\pi T^3} \frac{e^{\epsilon/T}}{(1+e^{\epsilon/T})^2}  d\epsilon.
 \end{eqnarray}
 Generalizing it, we obtain the formula,
 \begin{eqnarray}
\frac{\kappa_{xy}}{T} = \sum_{n} \nu_{n}\big( \frac{\pi}{12} -\frac{1}{2\pi T^3}\int^{\infty}_{\Delta_{n}} \frac{\epsilon^2 e^{\epsilon/T}}{(1+e^{\epsilon/T})^2} d \epsilon \big),  
\end{eqnarray}
and we illustrate $\kappa_{xy}/T$ for $\vec{\nu}=(1,1,0,-1)$ and $\vec{\Delta} = (2,1,1,0.1)$ with a proper unit in Fig.~\ref{fig:kappa}.

\begin{figure}[tbp]
	\includegraphics[width=0.99\columnwidth]{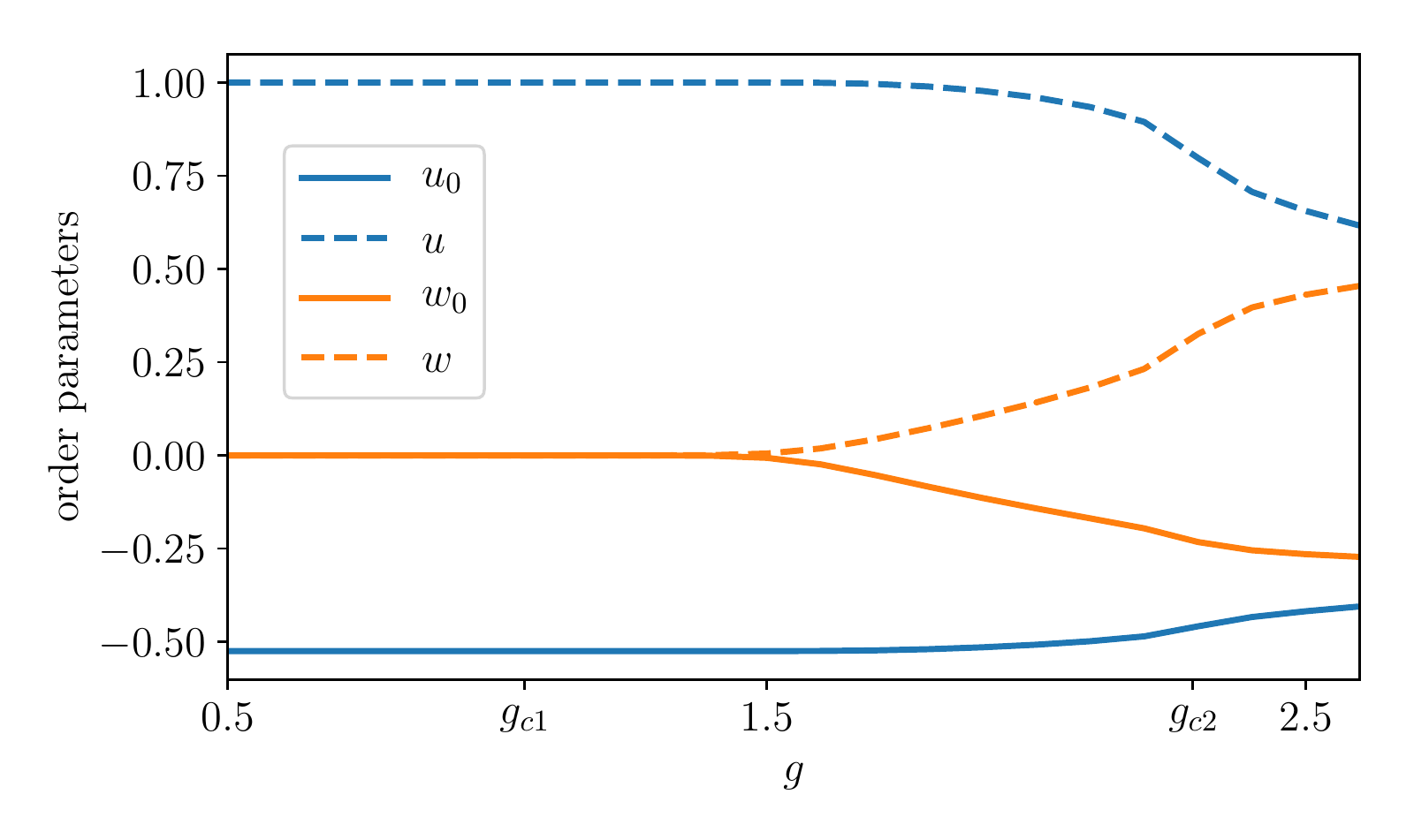}
	\caption{  Mean-field order parameters converged to nonzero values across the phase transitions at $g_\mathrm{c1}$ (from gapless to $\nu=1$) and $g_\mathrm{c2}$ (from $\nu=1$ to $\nu=2$).
	}
	\label{fig:order}
\end{figure}

\section{Mean-field order parameters}

The order parameters obtained by self-consistent mean-field calculations are given in Fig.~\ref{fig:order}.
As described in the main text, the nearest-neighbor interactions are decoupled by using the order parameters $(u_0, u)$ while the next-nearest-neighbor ones are associated with $(w_0, w)$.
In the gapless phase ($g<g_\mathrm{c1}$), the original Kitaev solution ($u_0=-0.5249$, $u = 1$) \cite{Kitaev} is reproduced.
The order parameters $(w_0, w)$ converge to nonzero values where the Chern number is well-defined as $\nu=1$.

\section{Berry curvature and Chern number}
 	The Berry curvature of $n$th band of the mean-field Hamiltonian at momentum $\mathbf{k}$ is computed from the Bloch function $u_{n\mathbf{k}}$ as Ref.~\onlinecite{Xiao2010}
\begin{align}
	F_n(\mathbf{k}) = \frac{\partial}{\partial k_x} A_{k_y} (\mathbf{k}) - \frac{\partial}{\partial k_y} A_{k_x} (\mathbf{k}),
\end{align}
where the Berry connection $A_{k_\alpha} = \langle u_{n\mathbf{k}} | \frac{\partial}{\partial k_\alpha} | u_{n\mathbf{k}} \rangle$ with $\alpha=x,y$.
Then the Chern number of the band $\nu_n$ is given by integration over the Brillouin zone as
\begin{align}
	\nu_n = \frac{1}{2\pi i } \int_{\mathrm{B.Z}} d \mathbf{k} F_n(\mathbf{k}).
\end{align}

\begin{figure*}[tbph]
	\includegraphics[width=0.90\textwidth]{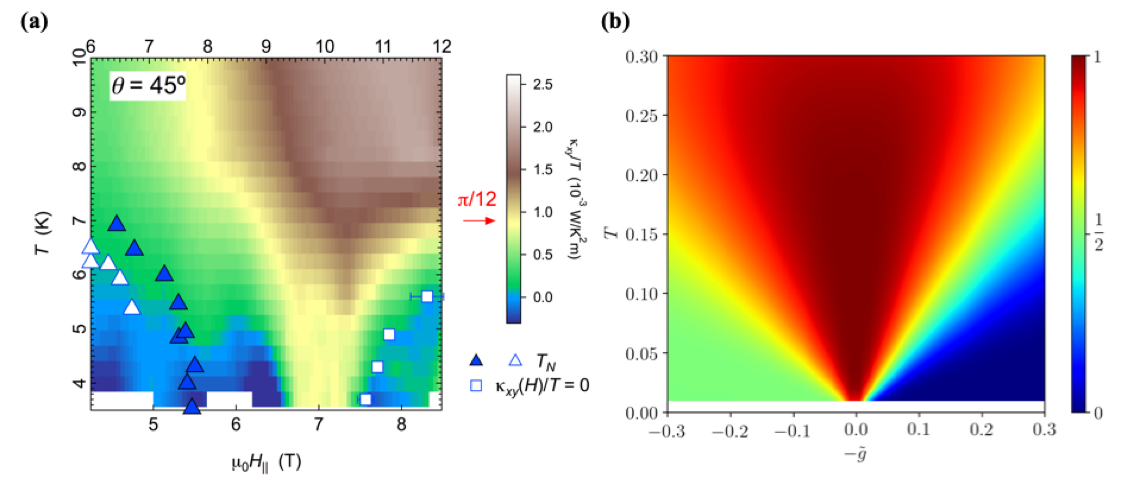}
	\caption{  Comparison  between (a) unpublished data of thermal Hall experiments, included with permission from Y. Matsuda, Kyoto University, and
		(b) theoretical calculation of $\kappa_{xy}^{edge}/T$. For comparison, we reverse the horizontal axis
	of the theoretical data with respect to Fig. 1(a) in the main text.
	}
	\label{fig:exp}
\end{figure*}

\section{Application to \texorpdfstring{$\alpha$-RuCl$_3$}{aRuCl3}}
In this section, we discuss a possible scenario of $\alpha$-RuCl$_3$ in terms of a TQPT starting from the original Kitaev model.  
In experiments, the TQPT is suggested at high magnetic fields based on the change of the Chern numbers from $\nu=1$ to $\nu=0$. 

We assume the three conditions. 
First, the phenomenological parameter dependence is $\tilde{g} \propto - B_{\parallel}$ near the TQPT. Second, the band gaps and Chern numbers have the same structures as ones of Fig. 1(a). The Chern number vector changes from $\vec{\nu}=(1,1,0,-1)$ to $(1,1,0,-2)$, and the smallest gap around the TQPT is associated with the $\nu_4=-1$ band.
Third, the $\nu=1$ phase ($\tilde{g}>0$) in Fig. 1(a) is connected to the original Kitaev model with a small magnetic field. 
The perturbative calculation and parton mean field analysis shows that the  Chern number vector $\vec{\nu}=(1,0,-1,1)$, and the smallest gap is associated with the $\nu_4=1$ band.
In Fig.~\ref{fig:exp}, we compare experimental data from the Kyoto group with our theoretical calculation.

We propose {\it a band crossing} to connect the $\nu=1$ phase in Fig. 1(a) with one of the Kitaev model under a weak magnetic field. 
The crossing is inevitable because the lowest energy band changes the Chern number from $\nu_4=1$ to $\nu_4=-1$. There are no changes in symmetry and topology, but the band crossing signatures would be observable energetically. 
Note that, in reality, the magnetically ordered state exists, so one should keep in mind that the band crossing point may be masked by the ordered phase.

\bibliographystyle{apsrev4-1}
\bibliography{Kitaev}